\documentclass[apjfonts]{emulateapj}

\shorttitle{LMXBs and Metallicity Dependence}
\shortauthors{N.\ Ivanova}

\begin{document}

\title{Low Mass X-ray Binaries and Metallicity Dependence: \\ Story of Failures}
 
\author{Natalia Ivanova$^1$}
 
\altaffiltext{1}{Northwestern University, Dept of Physics and Astronomy,
2145 Sheridan Rd, Evanston, IL 60208, USA.}

\email{nata@northwestern.edu}

\newpage

\begin{abstract} {Observations of galactic and extra-galactic globular clusters have shown 
that on average metal-rich clusters are $\sim 3$ times as likely to 
contain a bright X-ray source than their metal-poor counterparts.
We propose that this can be explained by taking into account 
the difference in the stellar structure of  main sequence  donors with masses 
between $\sim 0.85 M_\odot$ and $\sim 1.25 M_\odot$
at different metallicities.
Metal-poor main sequence stars in this mass range do not have 
an outer convective zone while metal-rich stars do.  
The absence of this zone turns off magnetic
braking, a powerful mechanism of orbital shrinkage, leading to the
failure of dynamically formed main sequence - neutron star binaries to
start mass transfer or appear as bright low-mass X-ray binaries.
}
\end{abstract}


\section{Introduction}

It was first noted for bright low-mass X-ray binaries (LMXBs) in globular clusters (GCs) 
of our Galaxy and of M31 that they preferentially reside in metal-rich GCs 
\citep{1993ASPC...48..156G,1995ApJ...439..687B}, where respectively 13 and 19 bright LMXBs 
with $L_{\rm X} \ga 10^{36}$ ergs s$^{-1}$ were observed
(in total, in both metal-rich and metal-poor clusters).
Recent extragalactic observations with a much larger sample of GCs and observed bright LMXBs 
have confirmed this tendency.
In particular, in NGC 4472 (Kundu et al. 2002), 30 bright LMXBs ($L_{\rm X} \ga 10^{37}$ ergs s$^{-1}$) 
were detected and identified with 820 GCs,
23 of these bright LMXBs are in the 450 metal-rich GCs and 7 -- in the 370 metal-poor GCs (Maccarone et al. 2004).
In M87, Jordan et al. (2004) have found 58 bright LMXBs ($L_{\rm X} \ga 10^{37}$ ergs s$^{-1}$) 
associated with 1688 GCs and have shown that
metal-rich GCs are $3\pm 1$ times as likely to contain bright LMXBs as metal-poor GCs.
As metal-rich clusters  we define here those with about solar metallicity, and
as metal-poor clusters those with metallicity $< 0.1$~Z$_\odot$.
The formation rate of bright LMXBs in GCs is about 100 times larger than in the field and it was
suggested (and now is generally accepted) that bright LMXBs in GCs are formed 
through dynamical encounters \citep{1975ApJ...199L.143C}.
Therefore, it seemed to be obvious  to expect that the preference of bright LMXBs for metal-rich clusters is due to 
some mechanism of the dynamical LMXB formation, which only works, or is more efficient, in metal-rich clusters.

Several attempts to explain the discrepancy were made.
Grindlay (1993) suggested that this may be due to a flatter initial mass function 
in higher metallicity GCs. To our knowledge, this has not yet been confirmed by observations,
and it is generally believed that the initial mass function is fairly universal (e.g., Kroupa 2002).
Secondly, it was proposed that dependence  of stellar radii on metallicity may be responsible.
\cite{1995ApJ...439..687B} 
suggested that in clusters of the same age, metal-rich stars are larger, 
and fill their Roche lobe  more easily, and are 
also more likely to experience binary formation via tidal captures. 
\cite{2004ApJ...606..430M} estimated that this gives only $\la 30-60\%$ 
difference in the formation rates. 
Another proposed idea is that in metal-poor clusters lifetimes of LMXBs are shortened  
by the extra mass loss caused by the irradiation-induced winds \citep{2004ApJ...606..430M}.
In this case the wind-mass loss is much higher than the mass-loss rate 
associated with the Roche lobe  overflow. 
Such strong winds may potentially be observable.

In this paper, instead of inventing a new mechanism of bright LMXB formation,
we explain why potential LMXB metal-poor binary systems fail to become bright LMXBs.
We review first how a close binary with a neutron star (NS) can be formed.
We analyze which of the dynamically formed binaries can achieve Roche lobe 
overflow.
We show how this ability to start mass transfer (MT) and to appear as a bright LMXB
hinges on the metallicity-dependent stellar structure of the donor, in particular
on the presence of the outer convective zone.
In Section 4 we estimate the formation rates and compare them with observations.

\section{Successful LMXBs Candidates}

There are several well-known ways to form a close binary with a NS: 
a physical collision with a Red Giant (RG), an exchange encounter, and a tidal capture.

In the case of a physical collision, a close binary with  white dwarf (WD) is formed \citep{1987ApJ...312L..23V}.
Binary evolution is driven here by gravitational radiation  only, and a significant fraction of such binaries 
can become Ultra-Compact X-ray Binaries (UCXBs).
The predicted formation rates for such binaries are in approximate accordance with 
observations \citep{2005ApJ...621L.109I}. However, these preliminary estimates, based on static cluster background 
and only considering single RGs, did not show that the UCXB formation 
rates are significantly  different in metal-poor and metal-rich clusters.
It is still possible that detailed  numerical simulations of GCs, including binaries, 
will reveal a difference.

From the available, although rather limited, set of observations 
for luminous LMXBs in galactic GCs, it can be seen that 
the two currently known GC-residing persistent LMXBs  
with a possibly  non-degenerate (but also not a subgiant)  companion are contained 
only in metal-rich clusters \citep[for data on LMXBs in GCs, see][]{BH_book_ch8};
the same is true for the two known GC-contained luminous transient, possibly  non-degenerate,  LMXBs.
On the other hand, among the five persistent LMXBs observed in relatively metal-poor clusters, 
four show indications of being ultracompact and one is likely to be a subgiant according to its orbital period.
Additionally, one luminous transient LMXB in a metal-poor cluster is identified as an ultracompact, and 
for the bright LMXB in Terzan 1 we have no information on whether it is ultracompact or not.
We note that only one of these five ultracompacts has its orbital period known and others
were identified as ultracompacts either by their X-ray to optical luminosity ratio 
or by interpreting their X-ray spectra.
Despite being limited, this data set nevertheless encourages us to test the hypothesis that
bright LMXBs that consist of a NS and a main sequence (MS) star can preferably be formed in
metal-rich clusters.
To do so, we will consider what range of binary periods, for a given eccentricity, can be formed 
via an exchange encounter or a tidal capture and will select only those binaries that can achieve MT while
the donor is a MS star.

\begin{figure}
{\plotone{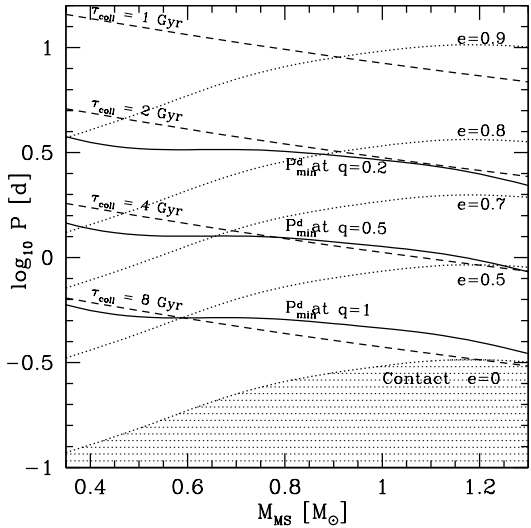}}
\caption{ Dynamically formed post-exchange (eccentric) binaries, 
MS star mass versus the binary period. The solid lines show minimum binary periods of post-exchange binaries 
$P_{\rm min}^{\rm d}$  assuming pre-exchange mass-ratios q=1, 0.5 and 0.2.
The dotted lines show binary periods at contact for corresponding eccentricities $P_{\rm min}^{\rm c}$.
The short-dashed lines show the collision times $\tau_{\rm coll}$. The collision times are found assuming   
1-d velocity dispersion of 5 km/s and the number density $n_{\rm c}=10^5$ objects per parsec$^3$.
For the shown period range, a dynamically formed post-exchange eccentric binary can have any period above 
both $P_{\rm min}^{\rm d}$ and $P_{\rm min}^{\rm c}$ for a given eccentricity.}
\label{nsms-exch}
\end{figure}

In an exchange encounter with a hard MS-WD or MS-MS binary, 
a NS replaces the smaller mass companion.
As the  binary energy in an exchange encounter is roughly conserved, the binary separation 
in the formed post-exchange binary $a$ is larger than the binary separation $a_0$ in the pre-exchange binary,
$a\sim a_{0} m_{\rm NS}/m_{2}$, where $m_2$ is the mass 
of the replaced companion \citep{1996ApJ...467..359H}.
Binary eccentricities in hard binaries  follow a thermal distribution, 
with an average eccentricity  $e\sim 0.7$ \citep{1975MNRAS.173..729H}
and post-exchange eccentricities in the case of an exchange to a more massive companion are also high
\citep{1996ApJ...467..359H}.
The minimum period of the post-exchange binary 
is limited by the period at contact for a given eccentricity $P_{\rm min}^{\rm c}$
and by $P_{\rm min}^{\rm d}$ defined by $a_{\rm min}^{\rm d}\sim a_{0}^{\rm c} m_{\rm NS}/m_{2}$,
where $a_{0}^{\rm c}$ corresponds to the pre-exchange binary separation at contact. 
The period at contact is defined here as the binary period at which the MS star overfills its {\it immediate} Roche lobe
at the pericenter. The applicability of this Roche lobe definition for eccentric binaries 
was shown in \citet{2005MNRAS.358..544R}. 
Dynamical formation of a binary with a period less than the period at contact is very unlikely --
such a binary will merge during an encounter \citep{2004MNRAS.352....1F}.
In Fig.~\ref{nsms-exch} we show minimum periods that dynamically formed eccentric binaries can possess.
The upper limit for the period of dynamically formed binaries will be at the boundary between soft and
hard binaries and is well above the periods shown on Fig.~1. 

We can estimate how long a post-exchange binary will remain
unperturbed by other stars in a GC during its subsequent evolution.
This is quantified by the collision time $\tau_{\rm coll}$, the time-scale 
for a binary to undergo a strong encounter with another
single star or a binary. 
For hard binaries, 
\begin{equation}
 \tau_{\rm coll}\approx \frac{\sigma}{4 \pi G a n } \left (m_1 + m_{\rm NS} + <m>\right)^{-1} \ ,
\end{equation}
where $n$ is the number density, $\sigma$ is the velocity dispersion and $<m>$ is the mass of another
single star or a binary.
E.g., in a typical dense cluster, a post-exchange binary with the orbital period more than a few days 
will likely undergo some kind of an encounter in its subsequent evolution 
and only a post-exchange binary with a period less than a day 
will likely remain intact (see Fig.~1).
The formation time-scale of these hard NS-MS binaries in the same cluster, per NS, is longer
than their collision time, 
\begin{equation}
\frac{\tau_{\rm exch}}{\tau_{\rm coll}} \approx \frac{m_{\rm NS}}{m_{2}} f_{\rm hb}^{-1} \ ,
\end{equation} 
where $f_{\rm hb}<f_{\rm b}$ is the binary fraction of hard binaries and $f_{\rm b}$ is the total binary fraction,
its value for GCs ranges from a few per cent to 40\% depending on the cluster density and is smaller in denser clusters
\citep{1996AJ....112..574C,2001ApJ...559.1060A,2002AJ....123.1509B,2005AJ....129.1934Z,2005MNRAS.358..572I}. 
Less than a few per cent of all NSs will have, within 10 Gyr (a typical GC age),
an exchange encounter with so hard a binary 
that the resulting post-exchange binary will have its orbital period less than a day.
Note also that the collision time for pre-exchange binaries 
is several times larger than $\tau_{\rm coll}$ of resulting binaries -- this suggests
that most of the pre-exchange binaries are primordial.

\begin{figure}
{\plotone{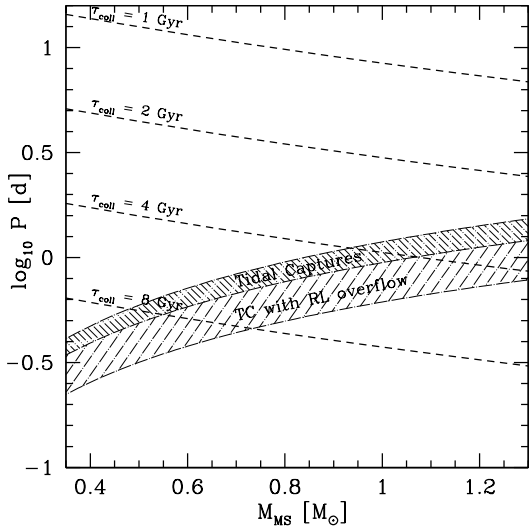}}
\caption{ Non-eccentric binaries formed via tidal captures (two dashed areas), MS star mass versus the binary period.
The lower dashed area shows binaries formed via tidal capture with a MS star overflowing its Roche 
lobe at the minimum approach.
The short-dashed lines show the collision times (see the caption for Fig.~1).
A binary formed via tidal capture can have a binary period only within the dashed areas.}
\label{nsms-tc}
\end{figure}

Close non-eccentric NS-MS binaries can be formed via tidal capture, where rapid circularization ($\sim 10$ years) is 
predicted by the standard model described in \citet{1987ApJ...318..261M}.
Using the approach described in 
\citet{Zwart_TC_93}, 
we can estimate  the post-capture binary parameters for a MS-NS binary (see Fig.~\ref{nsms-tc}),
where the upper limit corresponds to the closest approach at which tidal interactions
are still strong enough to make a bound system (this is found by equating the total kinetic energy
of NS and MS stars at infinity and the tidal energy that dissipates during 
the approach  -- see Eq~.(4) in {Portegies Zwart} \& {Meinen}, 1993).
For the lower limit we accepted a binary that would be formed if at the closest approach 
the radius of the MS star is 0.75 of its Roche lobe. 
We note that the parameter space for tidally captured binaries where MS star
does not overfill its Roche lobe at the closest approach is very restricted.
The collision time for tidally captured binaries is very long and their evolution is 
unlikely to be interrupted by other dynamical encounters after their formation.

Binary evolution of close NS-MS binaries  is driven by  angular  momentum losses  due to  magnetic
braking  and, to a lesser extent, gravitational radiation.
Stellar rotation also tends to synchronize with the orbit through tides.
In eccentric binaries, tides are also responsible for the eccentricity damping.
If tides are effective and reduce the eccentricity quickly, the binary is circularized at larger separations,
where gravitational radiation and magnetic braking
are less effective and therefore the consequent orbital shrinkage is slowed down.
We accepted for tides the prescription described in \citet{Hurley_Binary_02} and
for magnetic braking  the prescription based on X-ray observations of fast-rotating
dwarfs adopted from \citet{2003ApJ...599..516I}\footnote{Our main results, described later, do not depend on the
particular prescription of the magnetic braking law and will be valid for 
the Skumanich law based on the empirical relation for slowly rotating stars
from \citet{1983ApJ...275..713R}.}.

\begin{figure}
{\plotone{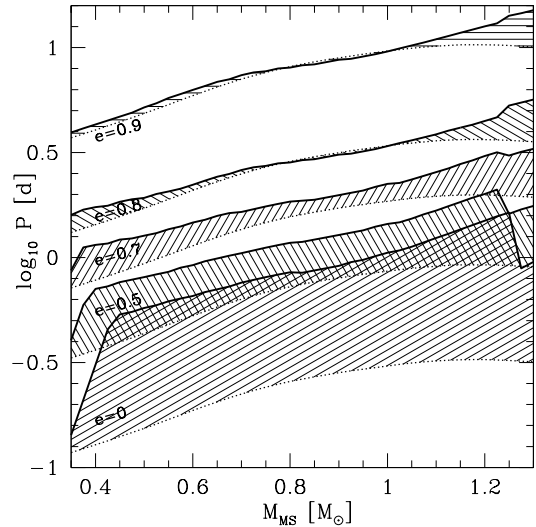}}
\caption{ Mass-transferring NS-MS candidates (MS star mass versus the binary period), metal-rich clusters.
The solid lines show the maximum binary period (for different initial eccentricities) such that the MS star
will overfill its Roche lobe during its MS life-time or before another dynamical encounter occurred.
Dashed areas show all possible initial binary periods that will lead to MT (for different initial eccentricities).}
\label{nsms-red}
\end{figure}

Taking into account the three processes described above 
(gravitational radiation, magnetic braking and tides), 
we can find the maximum initial binary period, for different eccentricities,
such that the binary will start the MT before the MS star leaves the MS 
or another dynamical encounter occurs (see Fig.~\ref{nsms-red}, where we show results for metal-rich clusters).
It can be seen that almost none of the highly eccentric ($e\ga0.75$) post-exchange binaries 
can start MT during their MS life-time. 
This is caused by the tidal eccentricity damping
(c.f. with the radiative stars with $M\ga 1.25 M_\odot$ -- a fraction
of them is able to evolve towards the contact).

The magnetic  braking efficiency depends  on the magnetic  field strength
and    the     density    of    the     stellar    wind    
\citep[see, e.g.,][]{1987MNRAS.226...57M}.  
According to observations of chromospherically
active binaries in GCs, it does not seem that magnetic fields in metal-poor clusters are qualitatively 
different from field (metal-rich) stars: close binaries with 
high rotation speeds show variability resembling starspots, consistent with that seen in
systems with similar rotation rates in the field \citep{2001ApJ...559.1060A}.
The wind  mass-loss rates in metal-poor
stars are  smaller than in metal-rich ones ($\propto Z^{1/2}$),  
and corresponding small reduction, about 30\%,   
of the magnetic braking efficiency in  metal-poor clusters can be expected.

\begin{figure}
{\plotone{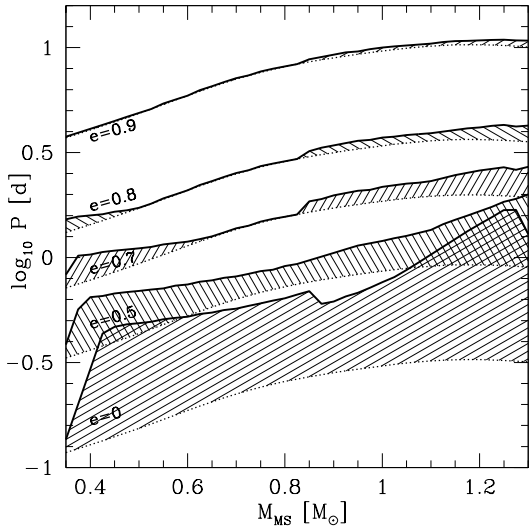}}
\caption{ Mass-transferring NS-MS candidates (MS star mass versus the binary period), metal-poor clusters.
Descriptions are as in Fig.~\ref{nsms-red}. The range of successful periods is decreased compared
to the case of metal-rich clusters (see Fig.~3).}
\label{nsms-blue}
\end{figure}

However the most important physical condition for magnetic braking  
to operate is the presence of a developed outer convective zone. 
From observations of stars of solar metallicities, only those with the developed
outer convective zone, $M \la 1.25 M_\odot$, 
show signs of high rate of angular momentum loss \citep[see, e.g.,][]{2003ApJ...582..358A}.
It is generally accepted that the mechanism 
of the magnetic braking is that the material lost from the 
stellar surface is kept in corotation with the star by 
the magnetic field. In the result the specific 
angular momentum carried by the gas is significantly greater 
than in a spherically symmetric stellar wind. 
To maintain corotation, a star should possess a substantial 
magnetic field and this can be achieved in low-mass MS stars 
by gas motions in the deep convective envelope, 
leading to the generation of a magnetic field by dynamo action
(for a short review how theoretical models of magnetic braking depend on 
the rotation and the magnetic field configurations see Ivanova \& Taam 2003).
A comparable outer convective zone  in metal-poor stars  is developed only in  stars $\la 0.85 M_\odot$.
Even more, as the convective zone in MS stars decreases with their evolution
through the MS, at the age of 10 Gyr only stars $\la 0.75 M_\odot$ have a developed outer convective zone.
As we lack the time-dependence for the moment of NS-MS binaries formation, 
we accept through this paper $0.85 M_\odot$ as a safe value for the border between 
the presence of the outer radiative and convective zones.
This simple  result can be  shown using  any available
single-star evolutionary  code; we obtained this using  the code described
in  \cite{2002ApJ...565.1107P} with the  updated  OPAL  opacities  for  low
metallicities (for low metallicity stars we adopted X=0.75 and Z=0.001)
and with the nuclear reactions rates taken from
Thielemann's library REACLIB  \cite{TTA}.
This  strongly  affects  how  many  dynamically  formed
binaries    can    achieve    the    Roche lobe   overflow    (see
Fig.~\ref{nsms-blue},   where   we   show   results   for   metal-poor
clusters).  Note  that  as  radiative  tides  are  not  as
efficient as tides in stars with  convective envelopes, 
a fraction of  highly-eccentric binaries can achieve the MT stage (c.f. metal-rich case).

\begin{figure}
{\plotone{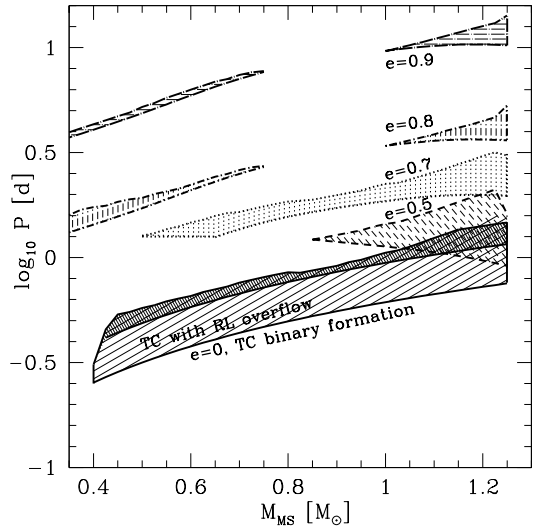}}
\caption{The population of NS-MS binaries that can be formed dynamically and can start MT,  metal-rich clusters.
This is the result of combining the limits described on Figs.~1, 2 and 3.}
\label{succ_red}
\end{figure}

\begin{figure}
{\plotone{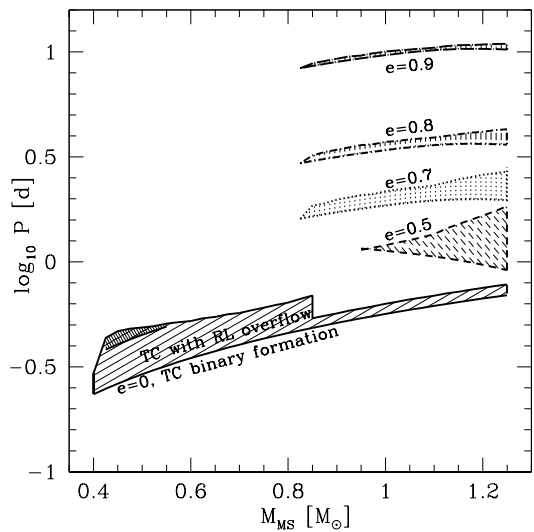}}
\caption{The population of NS-MS binaries that can be formed dynamically and can start MT, metal-poor clusters.
This is the result of combining the limits described on Figs.~1, 2 and 4.}
\label{succ_blue}
\end{figure}

Now we consider which dynamically formed candidate-binaries 
will successfully start the MT  in metal-rich and in metal-poor clusters.  
As an example, we consider the case with 
the mass ratio in the pre-exchange binary $q=0.5$
(this condition determines the minimum period of the post-exchange binary).
The particular choice of value of $q$ for our example is not significant,
other values of $q$ will give qualitatively similar results, but this value
resembles the observations well enough.
In particular, the observed distribution of $q$ for solar-type star at periods $\la 10^3$ d
is approximately flat \citep{1991A&A...248..485D}, and in very close
solar-type binaries, with the periods $\la 50$ days,  the mass ratios  $0.2\la q\la 0.6$ and $0.9 \la q\la 1$ 
are slightly more populated \citep{2004RMxAC..21...20H}.
The results are shown in Fig.~\ref{succ_red} and Fig.~\ref{succ_blue}.  
It can be seen that the the parameter-space available for post-exchange 
binaries that can successfully start MT in metal-rich clusters 
is substantially larger than in metal-poor clusters.  
Also, in metal-poor clusters  the efficiency of tidal captures is significantly reduced for 
MS stars with radiative envelopes ($\ga 0.85 M_\odot$) and only MS
stars with masses less than $0.55 M_\odot$ can be formed via tidal capture
without overfilling  their Roche lobe during the event. 

Let us consider in detail an example -- the case of a dynamically formed NS-MS
binary with a 1 $M_\odot$ MS star in metal-rich cluster.
Assume that the binary was formed via exchange encounter,  and pre-exchange companion
was a 0.5 $M_\odot$ star (q=0.5). When a NS replaces this companion, the minimum period of newly formed
binary is $\sim 1.1$ day, due to the approximate energy conservation. 
One of the consequences is that only highly eccentric post-exchange binaries 
(with $e\ga 0.6$)  can be  contact binaries (see Fig.~1); all binaries
with smaller eccentricities will have too large separations at the pericenter.
As a result, very narrow range of the post-exchange 
binary eccentricities, $0.5\la e\la 0.7$, as well as narrow ranges of post-exchange
(and accordingly pre-exchange) periods will lead to an LMXB formation.
If NS-MS binary was formed via tidal capture of a single 1 $M_\odot$ MS, 
the range of periastron distances leading to LMXB 
formation via tidal capture is about twice smaller than for the successful occurrence of the tidal capture --
not all successful tidally captured NS-MS systems will become LMXBs.
In a metal-poor cluster, the range of post-exchange 
binary eccentricities is even smaller, although in this case
more eccentric binaries are more likely to start a MT.
As the tidal capture does not occur, no post-encounter binaries with an
eccentricity $e\la 0.5$ can be both formed and start the MT.

\section{Persistent and Transient LMXBs}

The observed numbers of bright LMXBs and qLMXBs in Galactic GCs 
depend not only on the formation rate,
but also on the life-time of an LMXB in the corresponding stage (persistent or transient).
To examine NS-MS systems behavior,
we evolve binaries and obtain MT rates using the binary evolutionary 
code described in \citet{2004ApJ...601.1058I}, taking conservative MT.
With the  current knowledge of the origin of transient
behavior in XRBs \citep[for a recent review, see][]{BH_book_ch13},
if the MT rate is lower than the critical MT rate for a specific donor,
the accretion disk is expected to be thermally unstable
and the binary system is assumed to be a transient X-ray source.

\begin{figure}
{\plotone{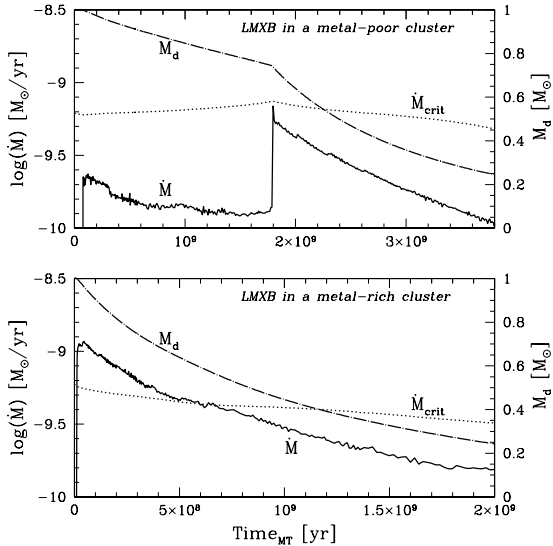}}
\caption{Evolution of NS-MS binaries, for the NS of 1.4 $M_\odot$ and the MS star with the initial mass of 1  $M_\odot$,
in a metal-poor cluster (the upper panel) and in metal-rich cluster (the bottom panel).
Shown are the MT rate $\dot M$ (solid line), the critical MT rate $\dot M_{\rm crit}$
(dotted line),  rates are in $M_\odot$ per yr.
The dash-dotted line shows the donor mass evolution.}
\label{lmxb}
\end{figure}

In metal-poor clusters, all binaries
with MS star $\ga 0.85 M_\odot$ (which are mainly post-exchange binaries), during
the initial stage of the MT, will evolve on the time-scale predicted by gravitational radiation.
The MT rates are significantly lower than in the case when magnetic braking is operating.
Also, in mass transferring MS donors the deep outer convective zone 
is developed at even smaller mass than 0.85 $M\odot$.
Comparing MT rates in our metal-poor LMXBs to the critical MT rate derived 
for hydrogen-rich donors from \citet{1999MNRAS.303..139D},
we find that all metal-poor LMXBs with a MS star $\ga 0.85 M_\odot$ (at the start of the MT) are transient
until they develop an outer convective zone (see Fig.~7). 
At this moment the MT rate sharply increases, but an LMXB stay transient.

In metal-rich clusters, an LMXB appears first as a persistent LMXB.
When donor decreases its mass to $\sim 0.7 M_\odot$, 
the MT rate becomes very close to critical and remains close for a long time. 
E.g., even when donor has already reached the mass of $\sim 0.25 M_\odot$, MT rate
is only 2 times smaller than the critical MT rate.
We note that the critical MT rate we use is not defined absolutely precisely, 
and carries uncertainties in our knowledge of accretion disks.
In the case of our LMXBs, even slight discrepancy in its value leads to large differences in 
how much time an LMXB spends as persistent $\tau_{\rm pers}$ or as a transient system $\tau_{\rm tr}$.
This uncertainty therefore affects how many qLMXBs and how many bright LMXBs can be present
in metal-rich clusters.
Also, it is possible that when the MT rate is about critical, a transient LMXB could have a greater
chance to be detected as a bright X-ray source, like MXB 1730-335 in Liller~1 that has
an outburst period comparable with the recurrence interval \citep{1996ARA&A..34..607T}.

The life-time of an qLMXB in a metal-poor cluster is longer than 
the life-time of a bright LMXB in metal-rich cluster.
Even if the number of created LMXBs in metal-poor cluster are several times smaller, 
the resulting number of LMXBs present at a given time can be such that 
a few qLMXBs will be present in metal-poor clusters per one bright LMXB in a metal-rich cluster,
if clusters are comparable by their dynamical characteristics. 
This estimate is rough and involves many uncertainties.
E.g., even the age of the MS star at the start of MT can affect drastically MT rates --
see, e.g., the discussion how MT rate varies with the donor age for the case 
of Skumanich-type magnetic braking law in King (2005).
It is therefore difficult to make a robust prediction on the ratio of qLMXBs
and bright LMXBs without having good statistics of the post-exchange 
NS-MS population and the corresponding study of MT episodes in all these systems.
However we may mention an interesting example of mildly metal-poor globular cluster 47~Tuc ($Z= 0.17 Z_\odot$): 
this cluster does not have a bright LMXB, but contains
at least two qLMXBs with a probably MS donors, X5 and W37 (orbital periods 8.7 and 3.1 hours) 
\citep{2003ApJ...588..452H,2005ApJ...622..556H}.

\section{Rates and Extragalactic Observations}
 
Let us examine now if the absence of the convective zone can indeed be
responsible for the observed ratio 3:1 of bright LMXBs in metal-rich 
and in metal-poor clusters.
This ratio would be easy to examine if it would reflect the formation rates 
in both types of clusters through the same mechanism only. 
However, we examine the idea that the dynamical formation of NS-MS bright 
LMXBs is almost suppressed in metal-poor clusters. Therefore this is a ratio
between the formation rates of UCXBs in metal-poor clusters and
combined formation rates of UCXBs and NS-MS bright LMXBs in metal-rich clusters,
and in both cases NS-RG LMXBs formation rates may play some role.
The theoretical predictions of the UCXBs formation rates are not robust
(see, e.g., Ivanova et al. 2005b), and none is available for
NS-RG LMXBs formation rates. At best we therefore can only estimate if
the theoretically predicted formation rates of NS-MS LMXBs can produce 
numbers of bright LMXBs similar to observations.

To estimate  the expected formation rate for NS-MS LMXBs we use 
available statistics for extragalactic clusters.
The detected bright GC LMXBs in other galaxies are not necessarily  
persistent sources, but also could be transient X-ray binaries in their outburst.
E.g., in M87, among GC X-ray sources with $L_{\rm X}>10^{38}$ ergs s$^{-1}$, about a third 
of sources show significant variability \citep{2004ApJ...613..279J}.
In GCs in our Galaxy, roughly a half of bright X-ray-source are transient \citep{BH_book_ch8}.
Additionally, the suggested ratio of qLMXBs and bright LMXBs is at least 7 to 1 \citep{2003ApJ...598..501H},
and may be even more \citep{2005ApJ...622..556H}.
But qLMXBs with a possibly non-degenerate donors in metal-poor clusters
have not yet been seen in outbursts. This implies that their duty cycles (the ratio of the time spent in the 
outburst to the recurrence interval) could be less than a few per cent, 
similar to many field NS X-ray transients \citep{1996ARA&A..34..607T}. 
E.g., if the duty cycle depends on the MT rate and becomes smaller when MT rates fall
significantly below the critical MT rate, it qualitatively correlates with
our finding for MT rates in metal-poor systems.
This suggests that bright transient metal-poor NS-MS LMXBs will unlikely 
contribute significantly to the set of observation for extragalactic GC LMXBs. 
In metal-rich clusters, the life-time of NS-MS LMXB in the transient stage
plays a significant role. 
This may introduce an error of a factor of 
$(\tau_{\rm pers}+\tau_{\rm tr}) / \tau_{\rm pers} \approx 4$ 
in the resulting rates.  
Please note however that in the case of metal-rich clusters, the duty cycle is also smaller than one,
and this reduces the possible number of transient source observed in the sample.

Let us consider NGC 4472, where, in the 450 metal-rich clusters,  23 bright LMXBs  were identified 
and in the 370 metal-poor GCs, bright 7 LMXBs were identified (Maccarone et al. 2004).  
We follow the steps described in \citet{2005ApJ...621L.109I} and find that, with the accepted 
retention factor for NS of 5\%, 450 metal-rich GCs contain  
$\sim 45000$  NSs in total, or $2000$ NS per one observed bright LMXB.
We suppose that one channel of bright LMXB formation works in both 
types of clusters and is the only formation channel in metal-poor clusters 
-- UCXBs formed through physical collisions.
This channel provides then 8-9 bright LMXBs in metal-rich clusters.
If we assume that all other bright LMXBs in metal-rich clusters are NS-MS binaries, then 
metal-rich clusters contain $\sim 3000-3200$ NSs per one observed bright NS-MS LMXB.
Accepting that the life-time of a bright NS-MS LMXB is $\tau_{\rm pers} \approx 0.3-0.5$ Gyr 
(c.f. $\tau_{\rm UCXB}\approx 10^7$ years), 
we find that the formation rate must be one bright NS-MS LMXB formation event per $\sim 90-160$ NS per 10 Gyr. 
An average GC has mass of $2.5\times 10^5\, M_\odot$ ($\sim 10^6$ stars) and
retains only about 200 NSs. Therefore, an average metal-rich GC produces only one or two bright NS-MS LMXBs 
during its entire life. 
This is just about the top of current abilities for numerical simulations of GCs 
(for Monte-Carlo type methods; $N$-body codes can deal with less than 1/10 this number of objects).
It is hopeless therefore to see this effect as the actual result of simulations. 

We may however attempt to estimate crudely the formation rate of bright NS-MS LMXBs in metal-rich clusters using 
the tidal capture formation rates.
The rate of tidal captures, per NS, in cluster cores can be found as: 

\begin{equation}
\tau_{\rm TC} = 80 \  {\rm Gyr }\ {\frac{v_{10}}{n_5} } \frac{1}{(p_{\rm max} - p_{\rm min}) 
(m + m_{\rm NS}) f_{\rm d}}\ ,
\end{equation}

\noindent where $v_{10}=\sigma/10 $ is the scaled cluster core velocity dispersion; 
$n_5=n_{\rm c}/10^{5}$ is the scaled cluster core number density; 
$p_{\rm max}$ and  $p_{\rm min}$ are maximum and minimum approaches that defined 
a successful tidal capture;
$m$ is the donor mass and $m_{\rm NS}$ is the mass of a NS; 
$ f_{\rm d}$ is the relative fraction of MS stars with masses between 0.8  and 1.25 $M_\odot$ (we take only donors
that produce persistent sources). 
The primordial fraction of these stars, in the case of Kroupa (2002) IMF, is $f_{\rm d}\approx 0.1$.
With time this fraction is slowly decreasing as MS turn-off mass decreasing, 
and at 10 Gyr $f_{\rm d}\approx 0.05$. 
Let us limit the consideration to the case of no Roche lobe overfilling at
the minimum approach and  $p_{\rm max}$ is found numerically by equating the tidally dissipated energy
and the total kinetic energy of NS and MS. 
The result can be written in an approximate 
form as $ (p_{\rm max} - p_{\rm min})/R_\odot\approx m_{\rm MS}/M_\odot-0.75$.
We find that the formation rate is one event per $\sim 130$ NS per 10 Gyr,
and is in good accordance with that estimated from observations.
 
We conclude that the discrepancy in bright LMXB formation rates is 
a natural consequence of the absence of the outer convective zones in 
MS stars of metal-poor GCs. Also, this effect has to be taken into account
when  modeling the population of different classes of
close binaries with a MS companion in metal-poor clusters.

\section*{Acknowledgments}

We thank C. Heinke, J.~Fregeau and B.\ Willems for helpful discussions, 
an anonymous referee for very useful comments that improved
the clarity of the manuscript and acknowledge support by a {\em Chandra} Theory grant.


\end{document}